\documentstyle[12pt,fleqn]{article}
\newcommand{\rb}[1]{\raisebox{1.5ex}[-1.5ex]{#1}}
\def\j{$J/\Psi $}
\def\pspr{$\Psi' $}
\def\J{J/\Psi\:}

\def\p0{$\pi^{0}$}

\def\be{\begin{equation}}
\def\ee{\end{equation}}
\def\ra{\rightarrow}
\def\P0{|\Psi_{0} \rangle\:}
\def\cc{c\={c}}

\def\lan{\langle}
\def\ran{\rangle}
\def\ovl{\overline}
\def\qgp{quark--gluon--plasma}
\def\cs{cross section}

\def\bd{\begin{displaymath}}
\def\ed{\end{displaymath}}
\def\xf{$x_{F}$}
\def\pt{$p_{\bot}$}

\def\xs{cross section}

\def\csq{$\chi^2$}
\def\csqp{$\chi^2_{\rm pdf}$}

\def\para{parametrization}
\def\distri{distribution}

\begin{document}

\title{\hskip 13cm {\large GSI-95-14} \\
       \vskip 1cm
    Empirical regularities in the $x$--dependence of nuclear \j \ suppression }

\author{C.Gerschel$^1$, J. H\"ufner$^2$ and E.Quack$^3$ \\
        $^1$ Institut de Physique Nucl\'eaire, 91406 Orsay Cedex, France \\
        $^2$Institut f\"ur Theoretische Physik der Universit\"at, \\
        Philosophenweg 19, 69120 Heidelberg, Germany \\
        $^3$ Theoret. Physik, Gesellschaft f\"ur Schwerionenforschung (GSI), \\
        P.O.Box 11 05 52, 64220 Darmstadt, Germany }
\date{January 1995}

\maketitle

\begin{abstract}
The measured ratios $R(x,A_1/A_2)$ for \j , $\Psi'$ and $\mu^+\mu^-$
production on two different targets $A_1, A_2$ as a function of the fractional
momentum $x$ of the final state are well fit with a
very simple functional form having three adjustable parameters.
An empirical relation is found between the three parameters of the fit.
The deduced \j \ absorption \xs s $\sigma_{abs}$
cluster around two values of which only $\sigma_{abs} = 5.8 \pm 0.2$ mb
seems acceptable.

\end{abstract}



With the advent of complete ${\cal O}(\alpha_s^3)$ calculations
\cite{nas,bee}, the production of charm states in hadronic collisions
like $hp \to \J X$ is approaching a quantitative understanding.
When produced in a nucleus, however, deviations
from the extrapolation of $hp$ reactions are observed \cite{na3} -- \cite{e789}
which are likely to be of a nonperturbative origin. Many
explanations have been proposed of which we quote only a few
\cite{kop} -- \cite{broho}. No definite conclusion on the mechanism
can be drawn at the present time.
The understanding of $hA$ collisions is of great importance as it
provides a reference for \j \ production in nucleus--nucleus collisions,
in connection with the search for \qgp \ formation.

While in the direction transverse to the beam an observed broadening of
the \pt \ distribution seems to be understood in terms of multiple
scattering of the initial partons \cite{pt}, the nature of the observed
longitudinal momentum dependence remains mysterious. In the present work,
we concentrate on the latter one and present an
analysis of the experimental data which is completely unbiased by
theoretical models. In this purely empirical way we find surprising
regularities which seem general to all the existing data.

Data are published using the kinematic variable $x_F = p_\|/p_\|^{max}$
in the hadron--nucleon center--of--mass system (cms),
and in this work we consider \pt --integrated data only. Mass $M$ and \xf \ of
the produced state can be related to the Bjorken variables $x_1$ and $x_2$ of
the partons in the projectile and target hadrons,
respectively, by $x_F = x_1 - x_2$ and $M^2 = x_1 x_2 s$ when
projectile and nucleon masses are neglected and where $\sqrt{s}$ is the
cms energy in the $hN$ system. Inverting these relations,
\be
  x_{1,2} = \frac{1}{2}[\pm x_F+\sqrt{x_F^2+4M^2/s}] \;
\ee
allows us to convert the observed values of $M^2$ and \xf \ into the
variables $x_1$ and $x_2$. M refers to the mass of the produced state $H$
($H$ = \j , $\Psi'$ or $\mu^+\mu^-$).

In the following, we study the ratio $R$ of the production \cs s on two
different target nuclei $A_1$ and $A_2$, where $x$ stands for one of the
variables
$x_1$, $x_2$ or $x_F$,
\be  R_{hH}(x,A_1/A_2) =
  \frac{ \frac{1}{A_1} \frac{d\sigma}{dx}(hA_1 \ra HX)}
  { \frac{1}{A_2} \frac{d\sigma}{dx}(hA_2 \ra HX)}
  \equiv A_1^{\alpha(x,A_1)-1} / A_2^{\alpha(x,A_2)-1} \; .
  \label{eq11}
\ee
We use the notation $R_{hH}(x,A)$ when dealing with a comparison of $hA$ with
$hp$ data ($A_1 = A, A_2 = 1$). In the frequently used parametrization
$A^{\alpha}$, $\alpha$ is the effectiveness of the nucleus $A$
to produce the heavy final state $H$ from the projectile $h$; a value of
$\alpha$ (or $R$) $< 1$ indicates a suppression.

While $\alpha = 1$ is expected from a perturbative calculation of heavy
final state production, experimentally a suppression is observed which
increases with \xf .
The data show relative small effects ($1-\alpha \le 0.02$)
for Drell--Yan dimuon pairs ($m_{\mu\mu} \geq $ 4 GeV) \cite{e772mu}.
The suppression is more sizable
for \j \ and $\Psi'$ production. Here, $1-\alpha \approx 0.05 - 0.09$ for
small values of \xf \ and increases to
$1-\alpha \approx 0.2$ at large \xf \cite{na3} -- \cite{e789}.

When experimental cross sections are concerned, the ratio eq.(2) is denoted
by $R_{hH}^{\rm ex}(x,A_1/A_2)$. Here, the
incoming hadron $h$ may be a proton, pion or antiproton, and the final state
$H$ a \j , \pspr \ or a $\mu\mu$ pair with given $M^2$.
The experimental cross section ratios $R^{\rm ex}(x,A_1/A_2)$ form the basis
of our analysis \cite{na3} -- \cite{e789}. We disregard several older
experiments with
very low statistics or a small acceptance range in \xf . Altogether, the
data come from 22 different reactions with a total
statistics of about $3\cdot 10^6$ events. The statistics differs strongly
from reaction to reaction, and
correspondingly do the errors on the parameter values which we will obtain.
We use the errors $\Delta R_{hH}^{\rm ex}(x,A_1/A_2)$ as they are given in the
papers, although it is not always clear to which degree they
include systematic uncertainties.

Several models have been proposed in order to explain the observed
\xf --dependence, among them parton energy loss in the initial state
\cite{kop},
shadowing of the parton \distri \ entering the hard fusion process \cite{gup},
final state absorption \cite{cj} and intrinsic charm in the projectile $h$
\cite{broho}. The assembly of these effects was considered in works such as
\cite{vog,kha}.
In a previous study \cite{cje}, we have analyzed the existing data in the
light of these theoretical models with the aim to discriminate between the
various proposed mechanisms. This attempt remained unsatisfactory
since the theoretical results did not describe the data well, i.e.~resulted
in an intolerable
high $\chi^2$, and the fits to different data sets were not always consistent.
In this paper we present a purely empirical study, void of any
theoretical prejudices, in which we try to find the best \para \ of the data.

We have analyzed experiments which differ in the projectile hadron, its
energy, in the target nuclei and in the final state (\j , \pspr \ and
$\mu\mu$). The data for the ratio $R(x,A)$ where $x = x_1$ or $x = x_F$
show two rather distinct regimes. For
low $x$, the ratio $R$ is approximately constant, while
it drops quite rapidly with $x$ at high $x$.
In order to investigate this property more quantitatively, we parametrize
the ratio $R(x,A)$ in the following form
\be
   R^{\rm fit}_{hH}(x,A) = A^{\alpha-1}\cdot [1-b\;(x-x_0)\Theta(x-x_0)]
 \label{eqfit}
\ee
with three adjustable parameters, the absorption coefficient $\alpha$, the
slope $b$ and the break point $x_0$. As usual, $\Theta(x) = 0$ for $x < 0$ and
$\Theta(x) = 1$ for $x \geq 0$. The functional form represented by
Eq.(\ref{eqfit}) consists of two straight lines intersecting at $x = x_0$: A
horizontal line for $x < x_0$ and a falling straight line with slope $b$ for
$x \geq x_0$.

By varying the parameters in Eq.(\ref{eqfit}), we perform
a least $\chi^2$ fit to the corresponding
data set, i.e.~we minimize the expression for \csq \ per degree of freedom,
\be
  \chi^2_{\rm pdf}(\alpha,b,x_0) = \frac{1}{N_{pts}-N_{par}}
  \sum_{i=1}^{N_{pts}}
  \left[ \frac{R^{\rm fit}(x^i,A_1/A_2;\alpha,b,x_0) -
   R^{\rm ex}(x^i,A_1/A_2)}
  {\Delta R^{\rm ex}(x^i,A_1/A_2)}\right]^2 \; ,
\ee
with respect to the parameters $\alpha,b$ and $x_0$. Here, $x$ may be
$x_1$ or $x_F$, the upper index $i$ runs over the $N_{pts}$ data points
measured in a particular reaction, and $N_{par}$ = 3.
The minimization is done numerically using
the routine MINUIT \cite{min}. The errors to the parameters
are also calculated with this routine and represent the
variance in each of the parameter values for a change of \csq \ from
the minimum value \csqp \ to $\chi^2_{\rm pdf} + 1$.

The results of this procedure are shown in Table 1 for
the data \cite{na3} -- \cite{e789} and for $x=x_1$.
The results for $x=x_F$ are not significantly different from these.
An ideal fit would give a value \csqp \ = 1.
The values for \csqp \ in Table 1 are mostly smaller than 1, which is
probably due to the inclusion of systematic errors into the quoted errors
of the data. The rather small values
of \csqp \ indicate that the \para \ by Eq.(\ref{eqfit}) is sufficiently
flexible to adjust well to the data.
Two representative fits are shown in Fig.~1.

Table 1 contains the values for the
parameters $\alpha$, $b$ and $x_0$, which
appear in the fit formula Eq.(\ref{eqfit}). The table also displays two other
parameters, the absorption \xs \ $\sigma_{abs}$ and the intercept $y_0$,
which are functions of the fit parameters.

The absorption \xs \ is an effective parameter which can be deduced from
$\alpha$ through the relation
\be
   A^{\alpha-1} \; = e^{-L \sigma_{abs} \rho_0}
 \label{eqabs}
\ee
where $\rho_0$ is the density of nuclear matter and $L$ the mean length
of the trajectory of the produced particle in the final state.
$\sigma_{abs}$ is an effective \xs \ as it describes the interaction with
nuclear matter of either the embryonic \cc \ \cite{kha} or of the final
\j \ or $\Psi'$.
Approximate formulae for $L$ are given by \cite{cj},
\be
  L = \left\{ \begin{array}{ll}
   \frac{A-1}{A} \frac{3}{4} r_0 A^{\frac{1}{3}} & A \ge 40 \\  &  \\
   \frac{A-1}{\rho_0} \frac{3}{8\pi} \frac{1}{\lan r^2 \ran } & A < 40 \\
 \end{array} \right.  \; .
 \label{eqabs3}
\ee
Here, $\lan r^2 \ran $ stands for the experimental mean square charged radius.
We use $r_0 = 1.2$ fm and $\rho_0^{-1} = 4\pi r_0^3/3$.

The intercept $y_0$ is defined as follows. One considers only the falling
straight line in Eq.(\ref{eqfit}) and denotes it by
\be
  y(x) = A^{\alpha-1}\cdot (1-b\;(x-x_0)) \; .
\ee
Then, extrapolating $y(x)$ to the point $x=0$ (where $y(x)$ should
not describe the data), one has
\be
  y_0 \equiv y(0) = A^{\alpha -1} \cdot (1+bx_0) \; ,
\ee
which depends on the parameters $\alpha$, $b$ and $x_0$. In Fig.~1, the
extrapolation $y(x)$ is shown by the dashed line.

We discuss the results shown in Table 1 and concentrate mainly on the \j \
data:
\begin{itemize}
\item[(i)] The values of the parameters do not display any systematic
correlation with the energy of the projectile. This phenomenon is called
scaling. The data are not sufficiently precise to distinguish between scaling
in $x_1$ or \xf . The reason for this difficulty is the following. The
suppression is constant, i.e.~independent of $x$ for small $x_1$ or \xf ,
where these variables differ. A significant $x$--dependence is observed only
above $x_0 \approx 0.3$, where $x_1 \approx x_F$. Therefore, $R(x,A)$ is
within the error bars not sensitive to the choice $x_1$ or $x_F$.
However, the observed scaling in $x_1$ or \xf \ implies
strong scaling violations in $x_2$, as has been pointed out already in
\cite{hoy} on the basis of fewer data.

\item[(ii)]
One cannot recognize any systematic variations in the parameter values which
correlate with the different projectiles $p, \pi$ and $\bar{p}$.
However, there are strong effects depending on whether the reference \xs \
in the denominator of Eq.(2) is measured on a proton ($A_2 = 1$) or on a
larger nucleus ($A_2 \geq 2$).

\item[(iii)]
The data seem to indicate
\be
  y_0 = 1.
\label{eqy0}
\ee
within the error bars. Of the 14 different experimental values 9 fulfill the
relation Eq.(\ref{eqy0}) within one standard deviation, while only two are
significantly outside the 2 standard deviation limit. Therefore we are rather
confident about the validity of the Eq.(\ref{eqy0}), which relates
the three fit parameters $\alpha, b$ and $x_0$ to each other.
The physical significance of the empirical result, Eq.(\ref{eqy0}), is
unclear to us. At present it indicates that the nuclear effects in the \j \
production from nuclei can basically be described by
{\em two} parameters only.

\item[(iv)]
The empirical values for the absorption parameter $1-\alpha$ or
equivalently the absorption \xs \ $\sigma_{abs}$ show
large fluctuations. A closer inspection reveals the following
systematics: The absorption \xs s deduced from the experiments where
nuclear data $hA$ are related to $hp$ ones with protons as targets cluster
around a value of
\be
   \ovl{\sigma}^{(1)}_{abs} = 3.2 \pm 0.5 {\rm mb}
\label{eqabs1}
\ee
where 4 out of the 5 experimental values are compatible with
$\ovl{\sigma}^{(1)}_{abs}$ within one standard deviation. The other data,
where the $hA_1$ results are related to $hA_2$ with $A_2$ = 2 or 9 cluster
around
\be
   \ovl{\sigma}^{(2)}_{abs} = 5.8 \pm 0.2 {\rm mb}
\label{eqabs2}
\ee
where 6 out of the 9 data points are compatible with
$\ovl{\sigma}^{(2)}_{abs}$ within one standard deviation.
\end{itemize}

The values $\ovl{\sigma}^{(1)}_{abs}$ and $\ovl{\sigma}^{(2)}_{abs}$ are
incompatible with each other within their error bars, thus the
classification into two groups may
indicate a physical effect: The two data sets deduced from $R(x,A/1)$
and $R(x,A_1/A_2)$, $A_2 > 1$ differ in their reference, one being the
production on a proton, the other on a small nucleus
having equal or nearly equal number of protons and neutrons.
The data with $A_2$ = 1 all come from the NA3 experiment \cite{na3}, while
the other experiments give only ratios with $A_2 \geq 2$. In order to
clarify the origin of the discrepancy in the two values for
$\sigma_{abs}$, we also use data for {\em absolute}
\xf --integrated \j \ production \xs s on various targets.

The absolute \xs s of \j \ production per nucleon
measured for $pp, pA$ and $A_1 A_2$ reactions at different projectile energies
are extrapolated to 200 GeV/nucleon by using the well established empirical
Craigie formula \cite{crai}
\[
 \sigma/{\rm nucleon} \propto {\rm exp}(\frac{-14.5M}{\sqrt{s}}) \; .
\]
These \xs s are given for $x_F > 0$ and $cos\theta_{CS}$ ranging from -1 to 1
where $cos\theta_{CS}$ is the Collins--Soper angle of the muon pair. When they
are plotted against the length $L$ of the trajectory in the final state,
they fall on one exponential curve with the exception of the $pp$ reaction
(NA3). This is shown in Fig.~2. Extrapolating the dashed line in Fig.~2 to
$L$ = 0 (no absorption), one deduces a
value of $\sigma = $ 4.4 nb for \j \ production in a $pN$ reaction,
where $N$ stands for the average of $pp$ and $pn$. This is off by two
standard deviations from the $pp$ result. Though unexpected, this result is
not in contradiction with fundamental symmetries, since $pp$ has isospin
I = 1 only, while I = 0 {\em and} I = 1 contribute to $pn$ reactions.
Yet the dominance of gluon fusion in \j \ production makes such large
differences in I = 1 and I = 0 appear very unlikely. Therefore we have to
conclude that the proton target data have a specific behavior in the NA3
experiment.

With the exception of the $pp$ point at $L=0$, all data shown in fig.~2 fall
on an exponential line whose slope is related to an absorption \xs \ of
$\sigma_{abs} = 5.9 \pm 1.4$ mb via Eq.(\ref{eqabs}). This value is
very close to the value $\sigma_{abs}^{(2)}$
extracted from the data shown in table 1. Only part of the experimental data
are in common between table 1 and fig.~2, as differential \xs s
$d\sigma/dx_F$ are not available for all systems.

Our discussion has shown that the $hp$ data fall out of the systematics
for $hA$ data when extrapolated to $A \to 1$.
Therefore we discard the value for $\ovl{\sigma}^{(1)}_{abs}$ as an
absorption \xs . Then the analysis based on Table 1 and Fig.~2 yield a value
\be
 \ovl{\sigma}^{J/\Psi N}_{abs} = 5.8 \pm 0.2 {\rm mb} \; .
\ee
As can be seen in Table 1, most of the deviations from the empirical law
occur for the data from NA3, which seem to have a problem with the
normalization. When correcting for the overall normalization of the NA3 data
by using the value of $\ovl{\sigma}^{J/\Psi N}_{abs}$, Eq.(12), the results
for $y_0$ are correspondingly shifted downwards and become compatible with the
observed systematics of $y_0$ = 1 from the other data sets.

Finally, we plot in Fig.~3 the ratios $R(x,A_1/A_2)/(A_1/A_2)^{\alpha-1}$,
where $\alpha$ is given in table 1. The normalization via the factor
$(A_1/A_2)^{\alpha-1}$ corrects for any possible discrepancies in the
absolute normalization of the $hp$ data. Only data are compared where $A_1$
is heavy (184 or 195) and $A_2$ is light (1,2 or 9). The solid line on the
figure is a fit to the $x_1$ dependence of the data using Eq.(3). The
similarity of the behavior of the different systems is striking despite
different projectiles and projectile energies and thus shows the degree of
scaling.

The data for $\Psi'$ production, measured only at 800 GeV, yield also an
absorption \xs \ of $\ovl{\sigma}^{\Psi'}_{abs} = 5.8\pm0.6$ mb which is the
same as for the \j \ final state. Since \j \ and $\Psi'$ differ in their
size and since the absorption \xs \ should depend on the radius, the
equality of the absorption \xs s is not obvious.
The explanation provided in ref.\cite{kha} is seducing but does not account
for all the regularities presented in this paper.
The systematics
$y_0 = 1$ is always fulfilled for the $\Psi'$ within the error bars.

The data for $\mu^+\mu^-$ production yield absorption \xs s compatible
with zero -- as it should be. The systematics of $y_0$ is unclear.

\vspace{0.5cm}

We summarize for the \j \ production on nuclei:
\begin{enumerate}
\item The data for the $x_1$ (or \xf )--distribution of \j \ production on
nuclei,
$R(x,A_1/A_2)$, can well be fit by a simple functional dependence of two
intersecting straight lines
\be
   R^{\rm fit}(x,A) = A^{\alpha-1}\cdot [1-b\;(x-x_0)\Theta(x-x_0)]
\ee
with three parameters $\alpha, b$ and $x_0$.

\item The parameters do not display any visible correlation with the type of
projectile nor its energy and thus show $x_1$ or \xf \ scaling.

\item The falling straight line of the \para \ when extrapolated to
$x=0$ takes the value $y_0 =1$ or
\be
  1 = A^{\alpha-1}\cdot (1+b\;x_0)
\ee
This empirical result seems well established, but its physical meaning is
not understood.

\item The absorption \xs s $\sigma_{abs}$ deduced from the fitted values for
$\alpha$ cluster around two distinctly different values
$\ovl{\sigma}^{(1)}_{abs} $ and $\ovl{\sigma}^{(2)}_{abs} $
depending on whether they are deduced from data for $R(x,A/1)$ or
$R(x,A_1/A_2)$, $A_2 > 1$, respectively. We argue that only
$\ovl{\sigma}^{(2)}_{abs} = 5.8 \pm 0.2$ mb can be interpreted as an
absorption \xs \ for \j \ propagation in nuclear matter.

\item The data for \j \ suppression in {\em nucleus--nucleus} collisions,
also displayed in Fig.~2, follow the same systematics as the $pA$ data, in
that a common absorption \xs \ of  $\sigma^{J/\Psi N}_{abs} = 5.9 \pm 1.4$
mb describes all data. This implies that so far no extra effects
are seen in nucleus--nucleus collisions which are not yet present in
proton--nucleus ones.

\end{enumerate}

\vspace{0.25cm}
{\em Acknowledgments} \\

We thank J.Dolejsi for his help with the minimization procedure. This work has
been supported in part by the BMFT under contract no. 06 HD 742.



\centerline{Figure Captions}

{\bf Fig.~1}\\
Left: The ratios $R(x,A)$ for $\pi Pt \ra \J$ relative to $\pi p \ra \J$ at
150 GeV from the NA3 experiment \cite{na3}, Right: $R(x,W/D)$ at 800 GeV
from the experiment E772 \cite{e772j}.
Both data sets are plotted as a function of $x_1$ and fitted by the \para \
eq.(\ref{eqfit}). The extrapolation of the falling straight line to the
intercept $y(0)$ is indicated by the dashed lines. \\

{\bf Fig.~2}\\
Plot of \j \ production \xs s/nucleon times branching ratio B($J/\Psi \to
\mu\mu$) for $pA$ and $A_1 A_2$ collisions as a
function of the mean length $L$ (in fm) of matter in the final state. All data
\cite{na3,e537}, \cite{ant} -- \cite{na38} are extrapolated to $\sqrt{s}$ =
19.4 GeV. The
exponential fit (dashed line) provides an absorption \xs \ of 5.9 mb. \\

{\bf Fig.~3}\\
Ratios $R(x,A_1/A_2)/(A_1/A_2)^{\alpha-1}$ plotted as a function of $x_1$.
Data
for $A_1$ = 184 and 195 and $A_2$ = 1,2,9 are fitted by the \para \ of Eq.(3).

{\small
\begin{table}[hbt]
\begin{center}
\begin{tabular}{|c|c|c|l||c|c|c|c||c|c|}\hline
\multicolumn{4}{|c||}{\rule[-3mm]{0mm}{8mm} \bf Reaction } &
\multicolumn{4}{c||}{\bf Parameters of fit } &
\multicolumn{2}{c|}{\bf Deduced quantities } \\ \hline
$H$ & $h$ & {\scriptsize $E_{Lab}$} & $A_1/A_2$ & $\alpha$ & $b$ & $x_0$ &
$\chi^2_{\rm pdf}$ & $\sigma_{abs}$ [mb] & $y_0$ \\ \hline\hline

   &  &200 &195/1$^a$ & 0.944 $\pm$  0.005&-2.4 $\pm$  0.6& 0.42 $\pm$
0.03&0.6 &  4.15 $\pm$  0.40& 1.51 $\pm$  0.20\\ \cline{3-10}
   &  &    & \ 12/2$^b$ & 0.952 $\pm$  0.018&-0.5 $\pm$  0.2& 0.25 $\pm$
0.09&0.4 &  5.02 $\pm$  1.84& 1.00 $\pm$  0.06\\ \cline{4-10}
   &  &    & \ 40/2$^b$ & 0.942 $\pm$  0.003&-4.1 $\pm$ 10.5& 0.57 $\pm$
0.51&0.4 &  5.50 $\pm$  0.27& 2.67 $\pm$  5.12\\ \cline{4-10}
   & $p$ &800 & \ 56/2$^b$ & 0.934 $\pm$  0.003&-1.3 $\pm$  0.7& 0.39 $\pm$
0.07&0.4 &  5.95 $\pm$  0.26& 1.16 $\pm$  0.23\\ \cline{4-10}
   &  &     &184/2$^b$ & 0.918 $\pm$  0.003&-1.2 $\pm$  0.3& 0.31 $\pm$
0.04&0.5 &  6.30 $\pm$  0.27& 0.90 $\pm$  0.07\\ \cline{4-10}
   &  &     &\  64/9$^c$ & 0.938 $\pm$  0.718&-0.4 $\pm$  0.6& 0.32 $\pm$
0.65&2.9 &  3.38 $\pm$ 39.46& 1.00 $\pm$  0.29\\ \cline{2-10}
   &  &     & 64/9$^d$ &  0.923 $\pm$  0.013& 0.7 $\pm$  0.5& 0.36 $\pm$
0.00&4.1 &  4.26 $\pm$  0.73& 0.65 $\pm$  0.14\\ \cline{4-10}
\rb{\j }  & \rb{$\ovl{p}$} & \rb{125} &184/9$^d$ & 0.885 $\pm$  0.007&-2.7
$\pm$  1.9& 0.59 $\pm$  0.06&0.6 &  6.08 $\pm$  0.35& 1.85 $\pm$  0.79\\
\cline{2-10}
   &  &    &64/9$^d$ &
0.990 $\pm$  0.559&-0.3 $\pm$  0.9& 0.22 $\pm$  0.53&4.2 &  0.57 $\pm$ 30.70
& 1.05 $\pm$  0.26\\ \cline{4-10}
   & \rb{$\pi^-$}  &\rb{125} &184/9$^d$ & 0.818 $\pm$  0.015&-1.1 $\pm$  0.2&
0.42 $\pm$  0.03&7.4 &  8.99 $\pm$  0.73& 1.21 $\pm$  0.08\\ \cline{2-10}
   & $\pi^-$ &150 & & 0.948 $\pm$  0.002&-1.0 $\pm$  0.0& 0.33 $\pm$  0.00&1.5
&  3.80 $\pm$  0.17& 1.00 $\pm$  0.01\\ \cline{2-3} \cline{5-10}
   &$\pi^-$  & & & 0.977 $\pm$  0.005&-1.4 $\pm$  0.4& 0.40 $\pm$  0.05&0.2 &
1.69 $\pm$  0.39& 1.39 $\pm$  0.16\\ \cline{2-2} \cline{5-10}
   & $\pi^+$ & \rb{200} &\rb{195/1$^a$} & 0.958 $\pm$  0.005&-0.7 $\pm$  0.4&
0.34 $\pm$  0.10&0.4 &  3.07 $\pm$  0.37& 1.00 $\pm$  0.13\\ \cline{2-3}
\cline{5-10}
 & $\pi^-$ &280 & & 0.960 $\pm$  0.003&-1.1 $\pm$  0.1& 0.32 $\pm$  0.02&0.9 &
2.91 $\pm$  0.21& 1.09 $\pm$  0.03\\ \hline \hline
 & & & \ 12/2$^b$ & 0.947 $\pm$  0.018&-0.9 $\pm$  0.6& 0.28 $\pm$  0.00&5.0 &
5.54 $\pm$  1.86& 1.10 $\pm$  0.15\\ \cline{4-10}
 & & & \ 40/2$^b$ & 0.950 $\pm$  0.011&-1.5 $\pm$  0.5& 0.20 $\pm$  0.08&1.0 &
4.69 $\pm$  1.07& 1.09 $\pm$  0.12\\ \cline{4-10}
\rb{$\Psi'$}    &\rb{ $p$} & \rb{800}  & \ 56/2$^b$ & 0.930 $\pm$  0.011&-1.5
$\pm$  0.8& 0.34 $\pm$  0.06&0.8 &  6.37 $\pm$  0.98& 1.14 $\pm$  0.20\\
\cline{4-10}
 & & &184/2$^b$ & 0.917 $\pm$  0.010&-1.0 $\pm$  1.3& 0.31 $\pm$  0.17&0.3 &
6.37 $\pm$  0.73& 0.85 $\pm$  0.29\\ \hline \hline
 & & & \ 12/2$^e$ & 1.002 $\pm$  0.003&-0.9 $\pm$  0.8& 0.64 $\pm$  0.06&0.9 &
-0.24 $\pm$  0.31& 1.55 $\pm$  0.55\\ \cline{4-10}
 & & & \ 40/2$^e$ & 1.011 $\pm$  0.122&-0.2 $\pm$  0.1& 0.15 $\pm$  0.54&0.5 &
-0.99 $\pm$ 11.43& 1.07 $\pm$  0.10\\ \cline{4-10}
\rb{$\mu\mu$}  & \rb{$p$} & \rb{800}  & \ 56/2$^e$ & 1.000 $\pm$  0.002&-0.1
$\pm$  0.1& 0.35 $\pm$  0.00&1.8 &  0.03 $\pm$  0.14& 1.05 $\pm$  0.02\\
\cline{4-10}
 & & &184/2$^e$ & 0.998 $\pm$  0.001&-0.7 $\pm$  0.3& 0.53 $\pm$  0.05&0.3 &
0.17 $\pm$  0.11& 1.33 $\pm$  0.18\\ \hline

\end{tabular}
\end{center}
\caption{}Result of the $\chi^2$ fit of the test function, Eq.(3), to the
experimental ratios $R_{hH}(x_1,A_1/A_2)$, Eq.(1). Data are of ref.
\cite{na3} for $^a$), \cite{e772j} for $^b$), \cite{e537} for $^c$),
\cite{e789} for $^d$) and \cite{e772mu} for $^e$).
Here, $H$ stands for the final state produced in
the reaction, $h$ for the projectile, and $A_1$ and $A_2$ denote the two
targets. The parameters of the fit are listed
in the middle columns, together with the corresponding $\chi^2_{\rm pdf}$,
and the deduced quantities (see text) are shown in the two last columns.
\end{table}}

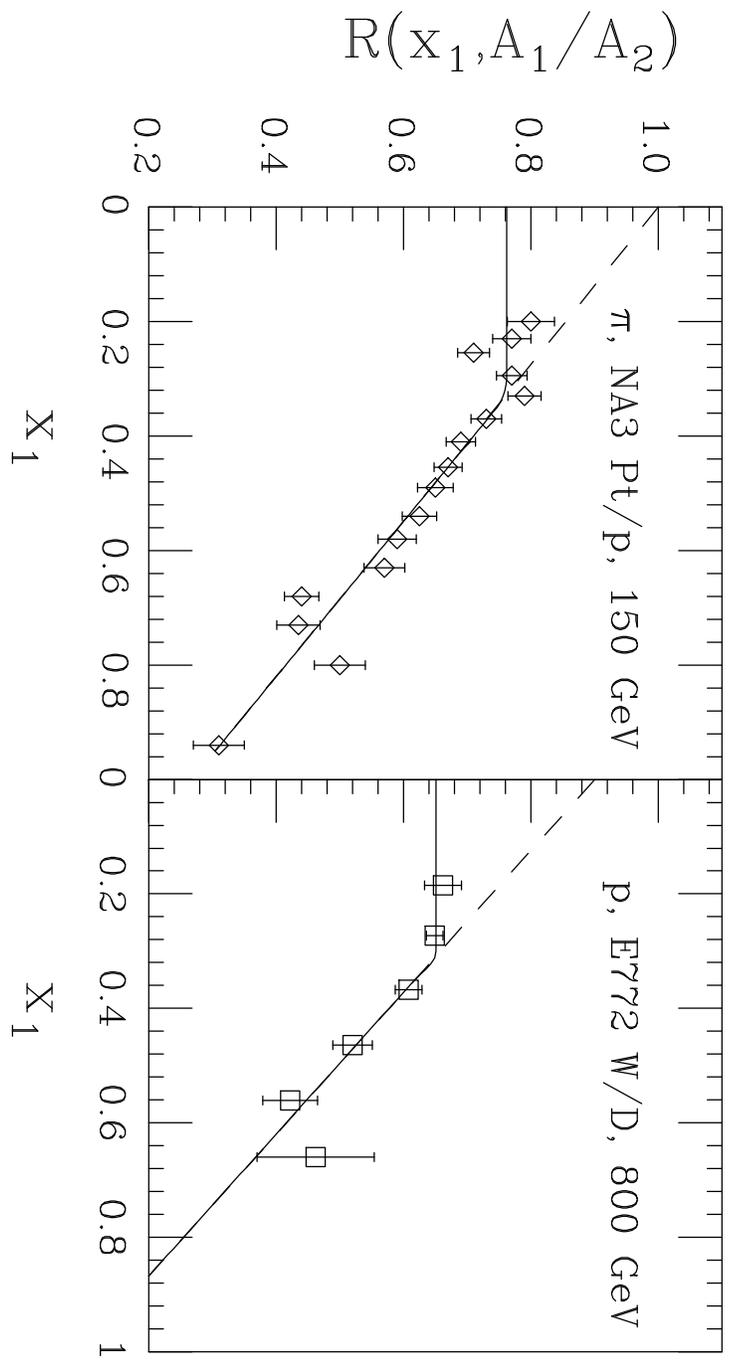
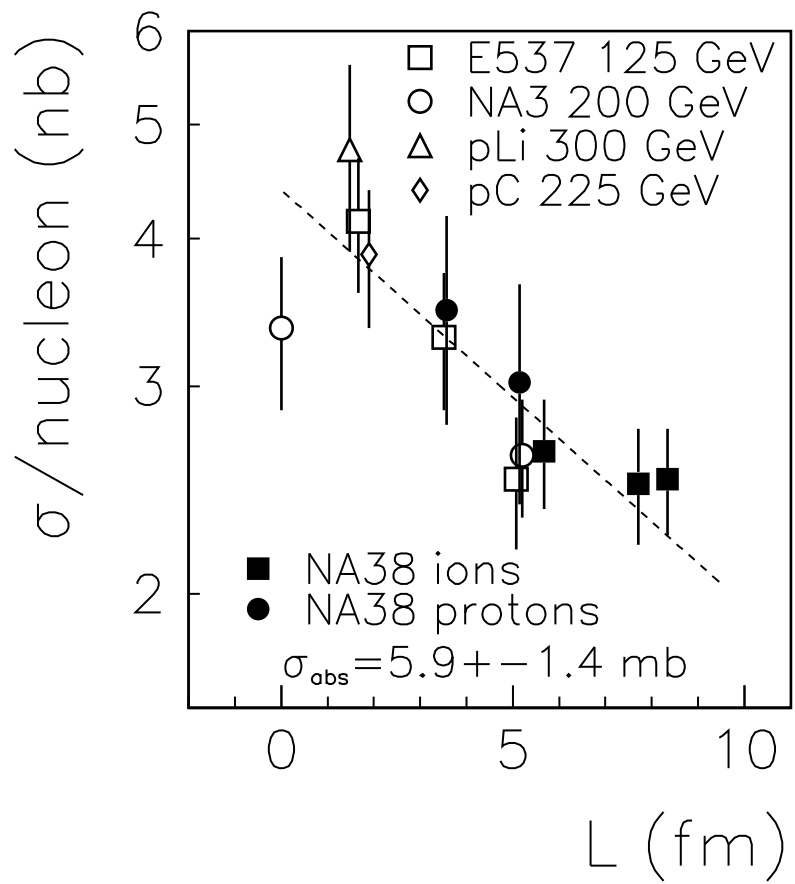
\end{document}